\documentstyle[12pt]{article}
\begin{document}
\normalsize
\renewcommand{\floatpagefraction}{ 2.}

\def \ih {\frac{{\rm i}}{\hbar}}
\def \QQ {{\bf Q}}
\def \HH {{\bf H}}
\def \ff {{\bf f}}
\def \cF {{\cal F}}
\def \Hb {{\bar H}}
\def \li {}
\def \bc {\begin{center}}
\def \ec {\end{center}}
\def \be {\begin{equation}}
\def \ee {\end{equation}}
\def \ben {\begin{displaymath}}
\def \een {\end{displaymath}}
\def \dd {{\rm d}}
\newcommand{\diffq}[2]{\frac{{\rm d} #1}{{\rm d} #2}}
\def \ii {{\rm i}}
\newcommand{\partq}[2]{\frac{\partial #1}{\partial #2}}
\def \ban {\begin{eqnarray*}}
\def \ean {\end{eqnarray*}}
\def \ba {\begin{eqnarray}}
\def \ea {\end{eqnarray}}
\mbox{}\\
\bc\Large\bf
THE THIRD WAY TO QUANTUM MECHANICS IS THE FORGOTTEN FIRST
\ec
\bc{\large\bf
by Salvatore ANTOCI\footnote{Dipartimento di Fisica ``Alessandro Volta'' dell' 
Universit\`a degli Studi di Pavia, Via Bassi 6, I-27100 Pavia, 
phone +39-382-507-486, fax +39-382-507-563}
and
Dierck-E. LIEBSCHER\footnote{Astrophysikalisches 
Institut Potsdam, An der Sternwarte 16, D-14482 
Potsdam,\\ email: deliebscher@aip.de, phone +49-331-7499-231, fax 
+49-331-7499-309}}
\ec
\bc
accepted by {\it Annales de la Fondation Louis de Broglie}
\ec
\vfill

\bc
\begin{minipage}{150mm}
\small
\bc
{\bf Abstract}
\ec
Quantum mechanics can be formulated in three ways, as Heisenberg, 
Schr\"odinger and Feynman did respectively. For the last way, an unknown 
(i.e. forgotten) forerunner exists, that we have found in a paper by 
Gregor Wentzel, published before the famous works by Heisenberg and 
Schr\"odinger, and contemporary with the fundamental works of L. de Broglie. 
In that paper, one can find the basic formulae and their interpretation 
as they were adopted by Feynman twenty years later. We believe that 
Wentzel's work was forgotten for several reasons: (I) Schr\"odinger's equation 
was much simpler to deal with (Wentzel himself contributed to its development 
in the same way as L.Brillouin and H.Kramers did). (II) The first application 
was rejected by Heisenberg and Kramers. (III) The approximation used by 
Wentzel was too na\"\i ve and failed. Nevertheless, the foundation laid 
by Wentzel was sound, as it has been shown by Feynman's work.
Therefore, Wentzel has to be considered as one of the 
founders of quantum mechanics.

Our exposition aims at explaining some details. It is accompanied 
by two appendices. They respectively provide a summary of the quoted 
paper by Wentzel and of the theory of canonical transformations, 
needed to understand the link between Wentzel's and Feynman's formulations.
\end{minipage}
\ec
\clearpage
\bc
\begin{minipage}{150mm}
\small
\bc
{\bf R\'esum\'e}
\ec
La m\'ecanique quantique peut \^etre introduit par trois chemins, celui de 
Heisenberg, celui de Schr\"odinger, et celui de Feynman. Pour le dernier, il 
existe un pr\'ecurseur inconnu, c'est-\`a-dire oubli\'e, que nous avons trouv\'e 
\'a un travail de Gregor Wentzel, publi\'e avant des travails c\'el\`ebres de
Heisenberg et Schr\"odinger, contemporain des travaux fondamentales 
de L.deBroglie. On y peut trouver les formules fondamentales et leur 
interpretation us\'ees par Feynman vingt ans apr\`es.
Nous croyons que le travail de Wentzel a \'et\'e oubli\'e par plusieurs raisons.
Premi\`erement, l'\'equation de Schr\"odinger \'etait plus simple \`a analyser 
(Wentzel m\^eme contribuait \`a ce developpement contemporainement avec 
L.Brillouin 
et H.Kramers). Secondement, l'application premi\`ere f\^ut r\'ejet\'ee par Heisenberg et 
Kramers. Troisi\`emement, l'approximation us\'ee par Wentzel \'etait 
trop simple et faillit. N\'eanmoins, la fondation par Wentzel \'etait correcte, 
comme les travaux de Feynman ont montr\'es. Par cons\'equence, il faut 
apprecier Wentzel comme un des fondateurs de la m\'ecanique quantique.

Notre exposition veut expliquer quelques d\'etails. Elle est accompagn\'ee par 
deux annexes, l'un resumant le travail cit\'e de Wentzel, l'autre la theorie des 
transformations canoniques necessaire pour comprendre la connection des formules 
de Wentzel avec celles de Feynman. 
\end{minipage}
\ec

\section{Introduction}

Traditionally, we teach about three ways to quantum mechanics,
attributed to Heisenberg \cite{HWE25,BJP25,BHJ26,DPA26}, 
Schr\"odinger \cite{SER26A,SER26B,SER26D,SER26E}, and
Feynman \cite{FRP48}.
\\

In the Heisenberg picture, 
Hilbert-space operators are substituted for the classical variables, and 
these operators may be represented by matrices. The classical
canonical equations of motion are translated by the
correspondence \cite{DPA26} between the Poisson bracket  
and the commutator into
\be\label{Heisenberg}
\diffq{\QQ}{t} = \ih (\HH \QQ - \QQ \HH)\ ,
\ee
\ben
{\bf Q}[t + \delta t] =  \exp[\ih {\bf H} \delta t] {\bf Q}[t] 
\exp[- \ih {\bf H} \delta t] \ .
\een
The state of the system in question is represented 
by a fixed Hilbert vector, which is not necessarily made explicit, if the 
representation is produced by the matrix elements themselves.
\\

The central issue of the Schr\"odinger approach 
is the equation of motion for 
a state Hilbert vector, whose representation in the function space $L^2$ 
is the wave function $\psi[x,t]$: 
\be\label{Schroedinger}
\ii\hbar \partq{\psi}{t} = {\bf H}\psi[x,t]\ , 
\ee
\ben
\psi[x,t + \delta t] = \exp[- \ih {\bf H} \delta t] \psi[x, t]\ .
\een
This Hilbert vector moves on the unit sphere in Hilbert space, 
\ben
\diffq{}{t} \int \psi^*[x,t] \psi[x,t] \dd x = 0\ ,
\een
and 
it is used to calculate the dis\-tribution of the measured classical 
variables. Eigenstates to the Hamilton operator characterize  
stationary states. The approach is equivalent to that of
Heisenberg \cite{SER26C}.
\\

The third method was published more than twenty years later by
Feynman \cite{FRP48,FHA65}. It aims at the immediate calculation of transition
probabilities, which later on can be translated into the
expressions for a corresponding wave function. In this picture,
transitions are given by a set of mediating 
trajectories in configuration or phase space, 
each contributing to the phase of a Hilbert
vector, whose squared amplitude is the desired probability. The
phase of an individual path is the classical action integral, and one 
obtains 
\ben
\psi[x,t+\delta t] = \int (x|x')_{\delta t} \psi[x',t] \dd x'
\een
with the transition contribution
\be\label{feynman}
(x|x')_{\delta t} =  K \exp[\ih \int_{x'}^x L[x,{\dot x},t] \dd t]\ .
\ee
The total amplitude is the sum of the
interfering contributions, formally 
\ben
(q_A|q_E) = \int {\cal D}q[t] \exp [\ih \int L[q^i,
{\dot q}^i, t] \dd t]\ .
\een
With a suitable choice of $K$, this relation is equivalent, 
in simple examples, to Schr\"odinger's equation (\cite{FRP48}, eq.18). 
The main point of Feynman's approach is the notion of 
interference of paths, not the mere equivalence of a mechanical path 
through configuration space with transversals to the 
surfaces\footnote{The character $S$ is used for both the solutions of 
the Hamilton-Jacobi equation $H[q,\partq{S}{q},t] + \partq{S}{t} = 0$ 
and for the action integral $S = \int_{x'}^x L[x,{\dot x},t] \dd t$ 
depending on the path. We will try to avoid confusion.}
of constant $S$. The mathematics of the integration over paths is a 
problem in its own right \cite{MCE51,GYA60,AHK76}.
\\

Feynman cites suggestions and remarks of Dirac
\cite{DPA33,DPA35,DPA45}, which are taken as hints to use a kind of
Huygens' principle to evaluate the evolution in time of the
quantum-mechanical wave function. He -- and all followers -- did
not cite nor recognize an early work by Gregor Wentzel
\cite{WGR24A}, where exactly the 
formulas written later by Feynman
are derived with the aim of obtaining the characteristics of wave propagation 
by techniques of point mechanics.
Wentzel's paper reached the editor of {\it Zeitschrift f\"ur
Physik} on February 2nd, 1924. It is the first paper which describes a 
method like Feynman's to construct transition probabilities. 
This method was not 
applied in full \cite{WGR24B,WGR24C}. Never\-theless,
the method was appropriate, and proven to be 
manageable by Feynman. Thus, Wentzel's paper should be acknowledged 
in the history of quantum mechanics.
Heisenberg's famous article on matrix mechanics dates from
July 29th, 1925 \cite{HWE25}, and is clearly second to Wentzel's. 
Schr\"odinger published about the
quantum wave equation from January 27th, 1926 on 
\cite{SER26A,SER26B,SER26D,SER26E}.   
Only DeBroglie wrote about matter waves already in september 
1923 \cite{DLO23A,DLO23B,DLO23C}, prior to G.Wentzel. 
\\
 
In the following, we first intend to sketch
the derivation used by Wentzel, and secondly, we will conjecture
about the question, why this article was never associated to the
development of Feynman's path integral approach. For the embedding of 
our particular topic into the  
history of quantum mechanics, 
we recommend the book by F.Hund \cite{HFR84}.

\section{The value of interfering paths}

Mechanics and geometrical optics are governed by integral
principles which attribute a value to each path in configuration
or phase space. The actual motions or propagations are
identified by local extrema of this value which we call action.
The simplest of these principles is Fermat's principle, where
the action is the integral over the refraction index. In point
mechanics, the integral in time of the Lagrange
function $L[q,{\dot q},t] = E_{\rm kinetic} - E_{\rm potential}$
is the general rule. For time-independent total energy $E$, it
contains the Maupertuis-Jacobi 
principle, which identifies the refraction index with
$\sqrt{E-E_{\rm potential}}$, thus providing the link  
between mechanics and geometrical optics. 
The name of Fermat is associated with the method of identifying the
rays with minima of  
the integral over the refraction index. 
The name of Hamilton is associated with the connection to wave
propagation. The surfaces  
of constant phase define transversal rays which are solutions to
Fermat's principle.  
The phase is proportional to some action. For wave phenomena,
the ray is produced by the interference implicit in Huygens'
principle, which excludes all other points of space in the limit of
infinitesimally small wavelength. This is the argument  
to understand the meaning of the extremum principle for the
action. Mechanics seems to fail 
only in explaining dis\-persion and interference phenomena.
\\

All this was known and used already in the dis\-cussion of early
quantum theory. The question then was how to dis\-tinguish between
pure paths of particles and true wave phenomena. In the realm of
geometrical optics this is not possible, and the history of this
recognition was deeply influenced by Einstein 
\cite{EAL21,EAL22}. Einstein constructs his results with
interfering paths too,  
but these paths are real paths from different source events. In
addition, the experiment 
which he proposed in \cite{EAL21} was wrongly evaluated and
ended in dis\-appointment.\\
  
In the historical evolution, quantum mechanics 
turned out to be wave mechanics. After the construction of the 
Schr\"odinger equation, i.e. the appropriate wave equation, the wave function
was interpreted as providing 
a probability phase \cite{BMA26A,BMA26B}, whose interferences
produced transition  
probabilities. Therefore, the 
notion used by Wentzel for reinstating the particle concept is to
interpret the result of interfering paths as transition
probability, i.e. to translate the language of wave theory into
the language of particle statistics. 
\\

Feynman defines the value of a path $q^i[t]$ through the
configuration space by the action integral $S = \int L[q^i,
{\dot q}^i, t] \dd t$. This action integral yields the
contribution of the path in question to the transition 
amplitude $(q_A|q_E)$ from the configuration $E$ to the
configuration $A$. This contribution differs from path to path
by a phase, and the total amplitude is the sum of the
interfering contributions.
\\

In the first paper of 1924 \cite{WGR24A}, Gregor Wentzel 
anticipated this idea in a strikingly
explicit fashion.  As we already noted, the dis\-cussion of the quantisation 
postulate at that time opened a broad acknowledgement of the correspondence 
between geometrical optics and point mechanics based on the 
variational principles of Fermat, Maupertuis and Jacobi 
\cite{BGR23,BLO24,CAH23,DWI23}. The main 
question was to implement wave characteristics, 
and the general proposition was to use Huygens principle
\cite{DCG23,DLO24A,EAL22,EEP24,SAD23}. This principle was usually formulated 
as integral theorem for the wave equation. The interference interpretation 
never involved the contributions of individual virtual paths.
\\

Wentzel is the first to consider the logical argument for the  
contribution of such paths in phase space to a
probability amplitude. The central issue is the measure of the deviation from 
the classical path. Wentzel chooses the integral $\int \sum_{i} Q^i \dd P_i$, in 
the canonical coordinates where the momenta $P_i$ are constants of 
motion (see appendix B).
In extended phase space, which can be constructed by (1) defining time as an additional 
coordinate $q^0$, (2) its conjugate momentum $p_0 = -W$ as minus 
the energy, the Hamiltonian $H[q,p,q_0] + p_0$
vanishes. Here, in any system of canonical coordinates
we obtain
\ben
L \dd t = \sum_{i=0}^n p_i \dd q^i = \sum_{i=0}^n p'_i \dd {q'}^i + \dd F
\een
and the variations $\delta S = - \delta S_1$ of the integrals 
\ben
S = \int_E^A \sum_{i=0}^n p_i \dd q^i\ ,\ \ S_1 = \int_E^A \sum_{i=0}^n q^i 
\dd p_i = \sum_{i=0}^n \left( (q^i p_i)|_{{\rm at} A} - (q^i p_i)|_{{\rm at} E} \right) - S\ , 
\een
are canonical invariants.
The phase $\phi$, at the moment identified with that yielding
the quantum-mechanical interference, is postulated by Wentzel 
to be the partly invariant\footnote{Canonical transformation 
possibly add a term not depending on the paths between given endpoints. 
Hence, the interference 
of the different contributions is not affected. The path-independent term 
corresponds to a phase factor in quantum mechanics. Wentzel's expression
differs form Feynman's by such a path-independent term.} measure
\be\label{phase}
\phi = - \frac{1}{h} \int_E^A \left( \sum_{i=1}^n q_i \dd p^i  - t\dd
W \right)
\ee
in these coordinates, for any path\footnote{The sign is corrected in Wentzel's 
second article.}. 
The quantum interference is supposed to be analogous 
to the wave interference. Wentzel writes: {\it Indem wir die 
klassische Wellenphase durch unsere Quantenphase ersetzen, 
ist es nun leicht, die wellentheoretische Interferenzformel in die 
Sprache der Quantenstatistik zu \"ubersetzen: Stehen dem Lichtquant 
verschiedene Wege $s$ von $E$ nach $A$ zur Verf\"ugung, so ist 
die Wahrscheinlichkeit, da\ss\ es auf einem beliebigen der Wege $s$ 
nach $A$ gelangt und dort absorbiert wird, nicht etwa gleich der 
Summe der Apriori-Wahrscheinlichkeiten der einzelnen Lichtwege $s$, 
sondern $J$ mal so gro\ss\footnote{By 
replacing the classical phase of the wave with our quantum phase,
it turns out to be simple to translate the interference formula  
of wave theory into the language of quantum statistics: If the
quantum of light  
may propagate along different paths from $E$ to $A$, the probability for 
going to $A$ along any of them and being absorbed there  
is not given by the sum 
of the a priori probabilities of the individual paths $s$, but $J$ times 
that value.}, wo}
\be\label{uebergansw}
J = \frac{(\sum\ff_s {\rm e}^{2\pi \ii \phi_s} )(\sum\ff_s^* {\rm
e}^{-2\pi \ii \phi_s})}{|\sum \ff_s|^2}\ .
\ee
Wentzel identifies the principle of 
zero deviation from ``mechanics'' with the Fermat-Jacobi
principle of least refraction-corrected path, as was general use in the 
dis\-cussion of the wave-particle duality at that time.
The decisive step however is to write an interference formula
for all different paths with the same start and the same end,
independent of their being ``mechanical'', i.e. solution of
the equation of motion, or not. Such an interference
formula did not exist  
before. This constitutes the difference to all other attempts to interpret
wave phenomena by particle motion at that time. 
In the particular situation described by
Wentzel, the probability is related to the amplitude of light 
($\ff_s$ is the vectorial amplitude of the classical wave), but
both the  
variables entering the formulae and the basic philosophy are not merely 
characteristic for
optics alone, but for mechanics in general. Wentzel
writes expressis verbis: {\it Die formale \"Ubereinstimmung des
Z\"ahlers mit dem Amplitudenquadrat superponierter Wellen
sichert dem Ansatz eine ausnahmslose
G\"ultigkeit, was die Beschreibung irgendwelcher
Interferenzph\"anomene anbelangt}\footnote{The formal coincidence of the
numerator with the square of the amplitude of superposed waves
ensures the ansatz to be universally valid
for the representation of interference phenomena of any kind.}.
We want to underline that 
the explicit use of the interference formula 
introduced by Wentzel is the 
decisive step to quantum particle mechanics. 
The same interference concept is the basic 
idea of Feynman's approach
in 1948 too. The gap to this approach consists in the explicit
technics for handling a path integral, i.e. for really
calculating a transition amplitude. 

\section{The response to Wentzel's article}
 
Before the publication, in 1925, of Heisenberg's method to calculate 
matrix elements, dis\-persion 
theory was one of the central topics of dis\-cussion. 
The question was how to get interference and dis\-persion with a flow
of particles (the quanta of light, named ``{\it Nadelstrahlung}'') 
\cite{EAL22,DWI23,LRF23,OBH23,SAD23,OBH24,SAD25,OBH25}. 
The classical theory of dis\-persion explained 
the relation between absorption lines and anomalous dis\-persion, 
but neither the number nor the 
narrowness of lines. 
Guth \cite{GEU29} characterized that time by the battle 
between wave and particle structure of light, the particle aspect always 
increasing its realm. Smekal \cite{SAD23} expected still a much
longer way to show that wave theory was not indis\-pensable for
optics\footnote{``{\it Bis zur Verwirklichung derartiger
Zukunftshoffnungen, welche in mancherlei 
Hinsicht geeignet waeren, das Dogma von der Unentbehrlichkeit 
wellentheoretischer Ueberlegungen in der Optik der Reflexion und Interferenz 
zu zerstoeren, ist aber vielleicht noch ein sehr weiter Weg.}'' 
(For the realization of such hopes for the future, which would 
be appropriate under several respects, namely, to destroy the 
dogma of the indis\-pensability of wave-theoretical arguments in 
the optics of reflection and interference, perhaps there is 
still a much longer way.)}. There had been several attempts to
outline a particle-based explanation of wave phenomena 
\cite{EAL22,DWI23,LRF23,OBH23,OBH24,OBH25,LGN26A,TSS26A,LGN26B,TSS26B}. 
None of them really reached the stage of a mathematically constructed 
theory, and none of them reached the stage of a basis 
for mechanics in general such as Wentzel's.
\\

Wentzel's paper was always understood to be part of that
dis\-cussion, even by Wentzel himself, and the by far ampler
importance of his postulates, for mechanics in general, went
along unnoticed. In addition, Wentzel used his scheme to argue
for the adoption of an intermediate orbit between initial and
final state of an atom interacting with light, and his results
were not backed by experiment. Kramers and Heisenberg used only
the initial orbit and could fit the data \cite{KHW25}. The
formulas which Wentzel developed for dis\-persion  
with his interference concept in mind \cite{WGR24C} were cited
by Kramers and Heisenberg \cite{KHW25}, but rejected in a
lengthy footnote\footnote{In short: ``{\it Es gibt keine
experimentellen Gr\"unde, die G\"ultigkeit einer einfacheren
Formel anzuzweifeln.}'' (There is no experimental motive to doubt
the validity of the simpler (old) formula.)}. The success of the
formula found by Kramers and Heisenberg and the change in
direction of the following evolution of quantum mechanics made
this judgement final. So Wentzel himself cites his paper only
once, in the fol\-lowing article ``{\it Zur Quan\-ten\-theo\-rie des
R\"ont\-gen\-brems\-spek\-trums}'' \cite{WGR24B}. Already there
the exposition of the method is banned to an
appendix\footnote{``{\it Zur Quantentheorie unperiodischer Systeme im
allgemeinen}''}. In the journal {\it Physikalische Berichte} we
find an abstract by Wentzel himself, which gives the impression
too, that he only marginally recognized the importance of his
approach for mechanics\footnote{{\it Es wird versucht, die
Interferenzerscheinungen vom Standpunkte der Lichtquanten aus
als fundamentale statistische Ph\"anomene zu verstehen. Die
M\"oglichkeit dazu ergibt sich daraus, da\ss\ die Lichtphase
$\int \dd s/\lambda$ durch die Bohrsche Frequenzgleichung
$hc/\lambda = \Delta W$ eine einfache mechanische Bedeutung
erh\"alt\ldots} (It is attempted to understand the interference
phenomena from the viewpoint of light quanta as fundamental
statistical phenomena. From here the possibility emerges that
the phase of the light obtains a simple mechanical meaning
through Bohr's frequency equation).}.
\\

Wentzel was quoted by his colleague in M\"unchen K.F.Herzfeld
\cite{HKF24}. It is not completely clear whether Herzfeld cites
Wentzel in reference to a program or to a theory in his
article\footnote{Herzfeld writes: {\it Die zweite Aufgabe
besteht \ldots in der quantentheoretischen Deutung des
Huygensschen Prinzips. Diese Aufgabe ist aber \ldots nicht
verschieden von der allgemeinen, welche die quantentheoretische
Deutung der Interferenz stellt. Sobald diese gel\"ost ist}
\cite{WGR24A}, {\it ist damit auch die Brechung usw. erkl\"art}
(The second task is the quantum interpretation of Huygens'
principle. This task is not dis\-tinct from the general one to
explain the interference in quantum theory. As soon as this
problem is solved, refraction etc. is explained too).}.
As far as we found out, Herzfeld \cite{WHK28} was the last one
to give a full account of Wentzel's approach, naming it
``Corpuscular theory of interference''. But new interest did not
arise. Citing Wentzel, Herzfeld mentioned also Beck
\cite{BGU27}. This shows that the interference notion seemed to
him to be more important than the integral over the manifold of
paths.  
\\

Other authors quoted Wentzel's concept occasionally. 
For instance, when Epstein and Ehrenfest \cite{EEP24} wrote
that coherence and interference would  
resist any attempt to understand, Smekal \cite{SAD24} answered
by citing Wentzel 
as an example for such an understanding.
A.Land\'e \cite{LAL25} too presumably felt the importance of the
concept, but he 
criticized it with an argument based on causality. However, this
argument would invalidate also  
Feynman, if correct and applicable.
Pauli \cite{WPA26} cites Wentzel and Herzfeld only together with
Ornstein and Burger  
\cite{OBH23,OBH24,OBH25},
which proves that he did not notice the far more general
importance and the explicitness  
of Wentzel's concept\footnote{``{\it Die Versuche von G.Wentzel,
K.F.Herzfeld und L.S.Ornstein u. H.C.Burger, 
die Ausbreitung des Lichtes in dis\-pergierenden 
Medien vom reinen Lichtquantenstandpunkt aus zu behandeln,
k\"onnen vorl\"aufig  
wohl noch kaum als befriedigend angesehen werden.}'' (The
attempts by G.Wentzel, K.F.Herzfeld and  
L.S.Ornstein and H.C.Burger to deal with the propagation of
light in dis\-persive  
media from a pure light-quantum viewpoint at present cannot yet
be considered satisfactory.)}. 
Pringsheim \cite{PPE29} and Kulenkampff \cite{KHE33} 
only quote the experimental argument against Wentzel's dis\-persion theory.
\\

Presumably under the influence of the successes of quantum
mechanics Wentzel changed his attitude
and restricted his work to the use of the classical
(``mechanical'') solutions and to the correspondence of the
Hamilton-Jacobi function to a single phase. 
In this way, Wentzel also found the
approximation scheme now known as 
Wentzel-Brillouin-Kramers method
\cite{WGR26B,BLO26,BLO26A,KHA26}. Here at last, the 
concept of interfering paths is
forgotten. All the studies concerning the 
correspondence between classical 
mechanics and quantum mechanics now concentrated 
on geometrical optics and 
canonical transformation theory \cite{JPA26A,JPA26B,VJH28}. 
The Schr\"odinger equation was fitted best to the task of
 calculating spectra and underlying energy levels, and 
both the problem of second quantization and of calculating 
transitions in more general problems were still ahead.
\\

Even in F.Hund \cite{HFR84}, who explicitely aims to answer 
the question whether quantum 
mechanics could have evolved differently, we 
find no hint to 
Wentzel's path integrals. 
\\

Dirac was the next to express the idea of getting probabilities
by superposition of paths \cite{DPA33}. Dirac connected it
mainly to the canonical tranformation theory in the direction of
the Wentzel-Brillouin-Kramers approximation, i.e. to geometrical
optics. The interference principle is formulated, but not
explicitly. The identification of a phase with the action
integral is the result of the construction of the unitary
evolution operator. This construction is not possible 
without the knowledge of quantum mechanics existing at that time 
in the Heisenberg or Schr\"odinger form. Feynman
\cite{FRP48} asserts that his work was inspired 
by Dirac's publications. Now Dirac was always very sparing of 
citations, so it is difficult to draw conclusions from his not
mentioning Wentzel. In the end, nobody recognized the outline
already formulated by Wentzel; even he himself apparently never came
back to the driving idea of his early work on quantum optics. 
\\

The reader of older literature is often suspected of falling into the
trap of reading things into a book instead of reading out of a book.
However, any book is complete only with the reader and changes
with its reader. At any time, literature is lost if not read
anew, with the knowledge and the capabilities added by the time
between, not only knowledge of more physics, but also capability
of deeper reading\footnote{J.L.Borges
writes in the essay {\it El libro} \cite{BJL79}: 
{\it 
Cada vez que leemos un libro, el libro ha cambiado, la connotaci\'on 
de las palabras es otra. Adem\'as, los libros est\'an cargados 
de pasado \ldots Si leemos un libro antiguo es como si ley\'eramos 
todo el tiempo que ha transcurrido desde el d\'\i a en que fue escrito y nosotros.
} (Every time when we read a book, the book has changed, the 
connotation of 
the words is different. In addition, the books are loaded with past \ldots
If we read an ancient book, it is as if we read all the time elapsed 
between  
the day when 
it was written and us.)}. 
\\

We want to thank H.-J.Treder for dis\-cussion and help with his private library,
and to acknowledge the contribution of L.Mihich (Pavia) and F.Antoci 
for help and hints with the citations.
\mbox{}\\
\mbox{}\\
\mbox{}\\
\mbox{}\\

\appendix
{\Large\bf Appendices}
\section{Wentzel's paper}

At the beginning of his article ``Zur Quantenoptik'' Wentzel observes 
that since Einstein's derivation of Planck's radiation law certain probabilities 
are attributed to the emission and absorption processes, but no more 
precise assertions are made. He intends to propose a general hypothesis 
for such probabilities, that in his opinion can help in overcoming
the contradiction existing in theoretical optics: wave theory of interference 
and polarization on one side, quantum theory of the spectral lines 
on the other side. To this end, he interprets the interferences 
as the offspring of underlying quantum-statistical laws.

In Section 1 of his paper Wentzel remembers that the most important foundation 
of the quantum theory is certainly the law that an atomic system cannot 
radiate if it finds itself in what 
he calls a mechanical state, i.e. a state in which the laws of 
classical mechanics are obeyed.Radiative processes are instead invariably 
associated with ``transitions'' for which the laws of classical 
mechanics do not hold. But not only the acts of emission and of absorption
are ``non-mechanical'', since the very presence of light propagating
through a transparent medium will cause non-mechanical perturbations in the atoms 
involved in the process.

In order to provide an invariant measure of the deviations of the 
intra-atomic motions from Hamiltonian mechanics, Wentzel considers the 
canonical coordinates $\beta_k$ and the conjugate momenta $\alpha_k$
associated with the atomic systems involved in the propagation of light. 
For simplicity, the $\alpha_k$ are assumed to be constant in the 
mechanical states. Then the desired measure is provided by the integral 
$\int \sum_k \beta_k\dd \alpha_k$. This integral is extended to 
the particular path in phase space that corresponds to the deviations 
from mechanics caused by a light quantum going from an emitting atom 
$E$ to an absorbing atom $A$ in a certain way.
Wentzel attributes to any path of this kind a phase
\be\label{A1}
\varphi = \frac{1}{h} \int \sum_k \beta_k\dd \alpha_k\ ,
\ee
where $h$ is Planck's constant. $\varphi$ provides the sought-after 
bridge between the quantum behaviour and the wave-like phenomena.

Wentel introduces the total energy $W$ of the atomic systems as one 
of the momenta ($\alpha_1$), and the time $t$ as the coordinate $\beta_1$ 
conjugated to it; the phase $\varphi$ is then defined 
as\footnote{Since Wentzel does not write the upper 
limit to the summation index $k$, it is possible to interpret this new 
definition in two ways: either it is the outcome of a 
canonical transformation performed in ordinary phase space, or it corresponds 
to an extension of the phase space itself by the addition 
of energy and time as a further 
conjugate pair. In the latter case the phase introduced by Wentzel is 
just the one considered by Feynman (apart for a wrong sign, 
and a path-independent term). In the appendix to a subsequent paper 
\cite{WGR24B} Wentzel clearly chooses the latter option, and also 
corrects the sign error of eq.(\ref{A2}).}
\be\label{A2}
\varphi = \frac{1}{h} (\int t\dd W + \int \sum_2 \beta_k\dd \alpha_k)\ .
\ee
As a check of his ideas within geometrical optics 
Wentzel envisages the simple system constituted by the atoms 
$E$ and $A$ exchanging a light quantum of energy $\Delta W$ 
that travels in vacuo with the velocity $c$ along the path 
joining the two atoms, and from his definition (\ref{A2}) of the 
phase, by retaining only the first addendum, he recovers 
Bohr's $
\Delta W = h\nu$ principle. Section 1 ends with the definition of the 
refractive index $n$, and with the remark that Fermat's principle 
$\delta\int n\dd s = 0$ can now be rewritten as 
$\delta\sum\int\beta_k\dd\alpha_k = 0$, i.e. as the requirement 
that for the rays of geometrical optics the integrated deviation 
from mechanics shall be a minimum.

In Section 2, Wentzel defines his interference formula. 
If the light quantum has several paths at his disposal for 
going from an emitting atom $E$ to an absorbing atom $A$, the overall 
probability of the process is not equal to the sum of the a priori 
probabilities associated with the individual paths, which is given 
by 
\be\label{A3}
\mid \cF_0\mid^2 = \mid\sum_s{\bf f}_s\mid^2\ ,
\ee
where ${\bf f}_s$ is the vector amplitude of the classical wave 
associated with the $s$-th path. The overall probability is 
instead supposed to be $J$ times the a priori probability, where
\be\label{A4}
J = \frac{(\cF\tilde{\cF})}{\mid \cF_0\mid^2}
\ee
and the complex amplitude $\cF$ is given by
\be\label{A5}
\cF = \sum_s {\bf f}_s \exp(2\pi \ii \varphi_s)
\ee
where $\varphi_s$ is the quantum phase defined by 
(\ref{A1}) or by (\ref{A2}). Wentzel emphasizes the general validity 
of his formula for interference processes of any kind, 
and the advantage of ensuring
a priori that the ``wavelength'' measured through the 
interferences and through the photoelectric effect are 
one and the same thing.

He further notices an essential feature: in his conception 
the emitting and the absorbing systems are intrinsically 
coupled. To him it is also noteworthy that no interference is conceivable 
without the presence of the absorbing system. Section 2 ends with a long Note 
dealing with the issue of the coherence length, as it can be confronted 
from the proposed viewpoint.

In Section 3 Wentzel outlines a theory of discrete spectra through a 
specialized use of his phase and interference formulae. While contemplating 
only the degrees of freedom of the emitting atom, he introduces 
the action variables $I_k$ and the conjugated angle variables
\be\label{A6}
w_k = t\cdot\partq{W}{I_k} + u_k\ ,
\ee
as it is customary when dealing with ``conditionally periodic'' systems. 
The $u_k$ appearing in (\ref{A6}) are undetermined phases, which are 
constant in the mechanical motions. Wentzel 
tentatively\footnote{Already in the subsequent paper \cite{WGR24B}
dealing with continuous spectra Wentzel changes his mind, and assumes that 
the emission and the absorption processes are characterized by the variation 
in time of both the action variables and of the undetermined phases.}
postulates that they remain constant also during the transitions, and 
assumes that, in order to get the transition probability, 
one shall modify the interference formula proposed in Section 2, since one 
shall not only sum the amplitudes associated with the individual paths, 
but also take the average over the undetermined phases $u_k$. Under 
these assumptions, he finds that the probability of transition can be 
nonvanishing only when the action variables change by an integer 
multiple of Planck's action quantum:
\be\label{A7}
\Delta I_k = n_kh\ .
\ee
Therefore, if initially ``quantized'', the atom will find itself 
after the transition in another quantized state.

Under the mentioned assumptions Wentzel calculates the expression 
for the amplitude $\cF$ and finds agreement with Bohr's 
correspondence principle for the intensity and for the 
polarization. Then he compares his result with the predictions 
of the classical wave theory, and asserts that through his theory 
one can describe refraction, reflection and double 
refraction just as it is done classically; in fact, he adds, 
Huygens' principle is just  based on interferences.

Wentzel ends the Section and the paper by observing that, while 
in the former quantum theory the action quantum $h$ had to be 
introduced twice, i.e. once in the $\Delta W = h\nu$ principle, 
and a second time in the quantum conditions, his theory 
allows to introduce it just once, in the expression (\ref{A1})
for the quantum phase.

\section{Canonical transformations}
Canonical coordinates $(q^i,p_k)$ are coordinates of the phase space. 
They are defined by the canonical form 
of the equations of motion, 
\ben
\diffq{q^i}{t} = \partq{H[q,p,t]}{p_i}\ ,\ \ \diffq{p_k}{t} = -\partq{H[q,p,t]}{q^k}\ .
\een
Because of the sign, we identify configuration coordinates $q^i$ and conjugate 
momenta $p_k$.
Canonical transformations are transitions from one set of canonical 
coordinates $(q,p)$ to another one ($Q,P$). The method of interest 
to generate canonical transformations is a Legendre transform. We assume
the existence of a generating function $F[q,P,t]$ depending on the old 
configuration coordinates $q^i$ and the new momenta $P_k$.
The function $F[q,P,t]$ generates the transformation by
\ben
Q^i = \partq{F[q,P,t]}{P_i}\ ,\ \ p_k = \partq{F[q,P,t]}{q^k}\ .
\een
The new Hamilton function is then given by
\ben
\Hb\dd t - \sum_{i=1}^n P_i\dd Q^i = H\dd t 
- \sum_{i=1}^n p_i\dd q^i + \dd(F - \sum_{i=1}^n Q^iP_i)
\een
or
\ben
\Hb = H + \partq{F}{t}\ .
\een
Any new Hamiltonian $\Hb[Q,P,t]$ can be constructed as long as 
the partial differential equation
\ben
\Hb[\partq{F}{P},P,t] = H[q,\partq{F}{q},t] + \partq{F[q,P,t]}{t}
\een
can be solved.
The Hamilton-Jacobi transformation aims at a vanishing new Hamiltonian and constant
configuration coordinates and momenta:
\ben
S = S[q,P,t]\ ,\ \ \Hb = H[q,\partq{S}{q},t] + \partq{S}{t} = 0\ ,\ \ 
p_i = \partq{S}{q^i}\ ,\ \ Q^i = \partq{S}{P_i}\ ,
\een
\ben
\Hb\dd t - \sum_{i=1}^n P_i\dd Q^i = H\dd t - \sum_{i=1}^n p_i\dd q^i 
+ \dd(S[q,P,t] - \sum_{i=1}^n Q^iP_i)\ .
\een
Then we get
\ben
\sum_{i=1}^n Q^i\dd P_i = - L\dd t + \dd S\ .
\een
By this transform, the congruence of paths in phase space can be 
mapped onto the initial values $Q^i, P_i (i=1,\ldots,n)$ of coordinates and momenta 
at some time $t_0$. 
If we characterize the states by ordinary coordinates $q[t]$ and
conjugate momenta $p[t]$, the integral 
\ben
\int\limits_E^A \sum\limits_{i=1}^n Q^i[q,p,t] \dd P_i[q,p,t]
\een
vanishes for the 
classical solution. The definition of the generating function yields 
\ban
\int\limits_E^A \sum\limits_{i=1}^n Q^i[q,p,t] \dd P_i[q,p,t] &=& 
- \int\limits_E^A L[q,{\dot q}, t] \dd t 
+ S[q_A,P_A,t_A] - S[q_E,P_E,t_E]\\
& = &
- \delta \int\limits_E^A L[q,{\dot q}, t] \dd t\ .
\ean
Because the left-hand side vanishes for the solutions of the
canonical equations (the ``mechanical'' paths in Wentzel's language), 
the right-hand side is not merely the action integral
corrected by the path-independent term $S_A-S_E$, but the
deviation in the action integral, which is zero for 
the extremal. 

If the Hamilton function does not depend on time explicitly, we may look for 
action-angle variables by separating the time from the Hamilton-Jacobi function. 
We obtain
\ben
H = H[q,p]\ ,\ \ \partq{H}{t} = 0\ ,\ \ S[q,P,t] = W[q,P] - E[P] t
\een
with the new transformation
\be\label{B1}
\Hb\dd t - \sum_{i=1}^n P_i\dd Q^i = H\dd t 
- \sum_{i=1}^n p_i\dd q^i + \dd(W[q,P] - \sum_{i=1}^n Q^iP_i)\ ,
\ee
\ben
p_i = \partq{W}{q^i}\ ,\ \ Q^i = \partq{W}{P_i}\ ,\ \ 
H[q,\partq{W}{q}] = E[P] = \Hb[P]\ .
\een
The canonical equation now determine the motion to follow
\ben
{\dot Q^i} = \partq{\Hb}{P_i} = {\rm const}\ ,\ \ 
{\dot P_i} = \partq{\Hb}{Q^i} = 0\ .
\een
One may interpret Wentzel's introduction of the energy as canonical momentum as 
follows. We solve $E = E[P]$ for
the first momentum $P_1$, and use $E$ instead of the old $P_1$
for the first new momentum, separating it from the others
($P_2, \ldots, P_n$). We then obtain
\ben
{\dot Q_1} = \partq{\bar H}{E} = 1\ \rightarrow \ Q_1 = t + t_0\ ,\ \
{\dot Q}_2,\ldots,{\dot Q}_n = 0\ .
\een 
and our special formula changes from (\ref{B1}) to
\ben
\Hb\dd t - \sum_{i=2}^n P_i\dd Q^i - E\dd t 
= -L\dd t + \dd (W - Et - \sum_{i=2}^n Q^iP_i)\ ,
\een
\ben
 \sum_{i=2}^n Q^i\dd P_i + t\dd E = -L\dd t + \dd W\ - E\dd t 
 = -{\bar L} \dd t + \dd W\ , 
\een
where ${\bar L} = \sum_1^n p_k{\dot q}^k$ is the reduced Lagrangian. 
This is Wentzel's formula in the ordinary phase space interpretation.

\normalsize

\newcommand{\bb}[8]
{
\bibitem{#2} {\sc #1} (#3): #7 {\it #4} {\bf #5}, #6.
}
\def \PR {Phys.Rev.}
\def \RMP {Rev.Mod.Phys.}
\def \PRS {Proc.Roy.Soc.}
\def \ZP {ZS f. Physik}
\def \CR {Comptes Rendus Acad.Sci.}
\def \JPR {J.de physique et le Radium}
\def \PNA {Proc.Nat.Acad.Sci.USA}
\def \SBerPAW {Sitzungsber.d.Preuss.Akad.Wiss.}
\def \Nw {Die Naturwissenschaften}
\def \AdP {Ann.d.Physik(Lpz.)}
\def \JMP {J.Math.Phys.}

\end{document}